\documentclass[12pt,a4paper]{article}
\usepackage{amsmath} 
\usepackage{amssymb}
\usepackage{graphicx,epsfig}
\usepackage[mathscr]{eucal}
\usepackage{color} 
\usepackage[numbers,sort&compress]{natbib}
\usepackage[english]{babel}

\usepackage{tabularx}

\usepackage{setspace}
\usepackage{xspace}

\usepackage{multirow}

\newcommand\as{\alpha_{\mathrm{S}}} 
\newcommand\f[2]{\frac{#1}{#2}} 
 
\def\to{\rightarrow} 
\def\nn{\nonumber}

\def\eps{\epsilon}

\newcommand\amp{\mathcal{M}}
\newcommand\ampF{\mathcal{M}_{\mathrm{fin}}}

\usepackage{scalefnt,ulem,pstricks}
\definecolor{blue}{rgb}{0., 0., 1.0}

\def\rcut{r_{\rm cut}}

\usepackage{array}
\newcolumntype{L}[1]{>{\raggedright\let\newline\\\arraybackslash\hspace{0pt}}m{#1}}
\newcolumntype{C}[1]{>{\centering\let\newline\\\arraybackslash\hspace{0pt}}m{#1}}
\newcolumntype{R}[1]{>{\raggedleft\let\newline\\\arraybackslash\hspace{0pt}}m{#1}}

\usepackage[colorlinks=true,allcolors={blue!70!black}]{hyperref}

\begin{document} 
\begin{titlepage}
\begin{flushright}
ZU-TH 10/21
\end{flushright}

\renewcommand{\thefootnote}{\fnsymbol{footnote}}
\vspace*{0.5cm}

\begin{center}
  {\Large \bf Mixed QCD--EW corrections to $\boldsymbol{pp\!\to\!\ell\nu_\ell\!+\!X}$ at the LHC}
\end{center}

\par \vspace{2mm}
\begin{center}
  {\bf Luca Buonocore${}^{(a)}$}, {\bf Massimiliano Grazzini${}^{(a)}$},\\[0.2cm]
  {\bf Stefan Kallweit${}^{(b)}$}, {\bf Chiara Savoini${}^{(a)}$} and {\bf Francesco Tramontano${}^{(c)}$}

\vspace{5mm}

${}^{(a)}$Physik Institut, Universit\"at Z\"urich, CH-8057 Z\"urich, Switzerland\\[0.25cm]

$^{(b)}$Dipartimento di Fisica, Universit\`{a} degli Studi di Milano-Bicocca and\\[0.1cm] INFN, Sezione di Milano-Bicocca,
I-20126, Milan, Italy\\[0.25cm]

${}^{(c)}$Dipartimento di Fisica, Universit\`a di Napoli Federico II and\\INFN, Sezione di Napoli, I-80126 Napoli, Italy

\vspace{5mm}

\end{center}

\par \vspace{2mm}
\begin{center} {\large \bf Abstract} 

\end{center}
\begin{quote}
\pretolerance 10000

We consider the hadroproduction of a massive charged lepton plus the corresponding neutrino through the Drell--Yan mechanism.
We present a new computation  of the mixed QCD--EW corrections to this process.
The cancellation of soft and collinear singularities is achieved by using a formulation of the $q_T$ subtraction formalism derived from the next-to-next-to-leading order QCD calculation for heavy-quark production.
For the first time, all the real and virtual contributions due to initial- and final-state radiation
are consistently included without any approximation, except for the finite part of the two-loop virtual correction, which is computed in the pole approximation and suitably improved through a reweighting procedure.
We demonstrate that our calculation is reliable in both on-shell and off-shell regions,
thereby providing the first prediction of the mixed QCD--EW corrections in the entire region of the lepton transverse momentum.
The computed corrections are in qualitative agreement with what we obtain in a factorised approach of QCD and EW corrections.
At large values of the lepton $p_T$, the mixed QCD--EW corrections are negative and increase in size, to about $-20\%$ with respect to the next-to-leading order QCD result at $p_T=500$\,GeV.
\end{quote}

\vspace*{\fill}
\begin{flushleft}
February 2021
\end{flushleft}
\end{titlepage}

\section{Introduction}
\label{sec:intro}

The production of lepton pairs through the Drell–-Yan~(DY) mechanism~\cite{Drell:1970wh} is the most {\it classic} hard-scattering process at hadron colliders.
It provides large production rates and offers clean experimental signatures, given the presence of at least one lepton with large transverse momentum in the final state.
Studies of the resonant production of $W$ and $Z$ bosons at the Tevatron and at the LHC
lead, for instance, to precise determinations of the $W$ mass~\cite{Group:2012gb,Aaboud:2017svj} and of the weak mixing angle \cite{Aaltonen:2018dxj,ATLAS:2018gqq}.
It is therefore essential to count on highly accurate theoretical predictions for the DY
cross sections and the associated kinematical distributions. This in turn implies the inclusion
of radiative corrections to sufficiently high perturbative orders in the strong and EW couplings $\as$ and $\alpha$.

The DY process was one of the first hadronic hard-scattering reactions for which radiative corrections were computed.
The pioneering calculations of the next-to-leading order~(NLO)~\cite{Altarelli:1979ub}
and next-to-next-to-leading order~(NNLO)~\cite{Hamberg:1990np,Harlander:2002wh} QCD corrections to the total cross section were followed by
(fully) differential NNLO computations including the leptonic decay of the vector boson~\cite{Anastasiou:2003yy,Anastasiou:2003ds,Melnikov:2006kv,Catani:2009sm,Catani:2010en}. 
The complete EW corrections for $W$ production have been computed in Refs.~\cite{Dittmaier:2001ay,Baur:2004ig,Zykunov:2006yb,Arbuzov:2005dd,CarloniCalame:2006zq}
and for $Z$ production in Refs.~\cite{Baur:2001ze,Zykunov:2005tc,CarloniCalame:2007cd,Arbuzov:2007db,Dittmaier:2009cr}.
Very recently, the next-to-next-to-next-to-leading order~(N$^3$LO) QCD radiative corrections to the inclusive production of
a virtual photon~\cite{Duhr:2020seh} and of a $W$ boson~\cite{Duhr:2020sdp} became available.

Since the high-precision determination of EW parameters requires control over the kinematical distributions at very high accuracy,
the next natural step is the evaluation of mixed QCD--EW corrections. In the community, significant efforts have recently been devoted to this topic.

The mixed QCD--QED corrections to the inclusive production of an on-shell $Z$ boson were obtained
in Ref.~\cite{deFlorian:2018wcj} through an abelianisation procedure from the NNLO QCD results~\cite{Hamberg:1990np,Harlander:2002wh}.
This calculation was extended to the fully differential level for off-shell $Z$ boson production and decay into a pair of neutrinos (i.e.\ without final-state radiation) in Ref.~\cite{Cieri:2020ikq}.
A similar calculation was carried out in Ref.~\cite{Delto:2019ewv} in an on-shell approximation for the $Z$ boson, but including the factorised NLO QCD corrections to $Z$ production and the NLO QED corrections to the leptonic $Z$ decay.
Complete ${\cal O}(\as\alpha)$ computations for the production of on-shell $Z$ and $W$ bosons have been presented
in Refs.~\cite{Bonciani:2020tvf} and~\cite{Behring:2020cqi}, respectively.
Beyond the on-shell approximation, the most important results have been obtained in the {\it pole} approximation (see Ref.~\cite{Denner:2019vbn} for a general discussion).
Such approximation is based on a systematic expansion of the cross section around the $W$ or $Z$ resonance, in order to split the radiative corrections
into well-defined, gauge-invariant contributions. Contrary to the {\it narrow-width} approximation, the pole approximation (PA) describes off-shell effects of the
massive vector boson around the resonance.
Such method has been applied in Refs.~\cite{Dittmaier:2014qza,Dittmaier:2015rxo} to evaluate the so-called {\it factorisable} contributions of {\it initial--final} and {\it final--final} type. In these calculations the {\it initial--initial} contributions were neglected, and the so-called {\it non-factorisable} corrections were shown to be very small.

Given the relevance of mixed QCD--EW corrections for precision studies of DY production, and for an accurate measurement of the $W$ mass \cite{CarloniCalame:2016ouw},
it is important to go beyond these approximations.
New-physics effects, in particular, could show up in the tails of kinematical distributions, where the PA is not expected to work.
A first step in this direction has been carried out in Ref.~\cite{Dittmaier:2020vra}, where complete
results for the ${\cal O}(n_F\as \alpha)$ contributions to the DY cross section were presented.

One of the missing ingredients in the complete ${\cal O}(\as\alpha)$ calculation is the corresponding two-loop virtual amplitude. 
The evaluation of the $2\to 2$ two-loop Feynman diagrams with internal masses is indeed at the frontier of current computational techniques.
Progress on the evaluation of the corresponding two-loop master integrals has been reported in Refs.~\cite{Bonciani:2016ypc,Heller:2019gkq,Hasan:2020vwn}. Very recently,
the computation of the complete two-loop amplitude for the neutral current dilepton production process was discussed in Ref.~\cite{Heller:2020owb}.
On the other hand, the required tree-level and one-loop amplitudes can nowadays be obtained with automated tools.

In this paper we focus on the charged-current DY production process
\begin{equation}
  \label{eq:proc}
pp\to \ell^+\nu_\ell+X
\end{equation}
and present a new calculation of the mixed QCD--EW corrections.
At variance with previous work~\cite{Dittmaier:2014qza,Dittmaier:2015rxo}, we use the PA only to evaluate the finite part of the genuine two-loop contribution.
To this purpose, we carry out the required on-shell projection
by using the $W$ boson two-loop form factor recently presented in Ref.~\cite{Behring:2020cqi}, and we further improve our approximation through a reweighting technique.
All the remaining real and virtual ${\cal O}(\as\alpha)$ contributions are evaluated without any approximation. The corresponding tree-level and one-loop scattering amplitudes are computed with {\sc Openloops}~\cite{Cascioli:2011va, Buccioni:2017yxi, Buccioni:2019sur} and {\sc Recola}~\cite{Actis:2016mpe,Denner:2017wsf}, finding complete agreement.
The required phase space generation and integration is carried out within the {\sc Matrix} framework~\cite{Grazzini:2017mhc}.
The core of {\sc Matrix} is the Monte Carlo program {\sc Munich}\footnote{{\sc Munich}, which is the abbreviation of “MUlti-chaNnel Integrator at Swiss (CH) precision”, is an automated parton-level NLO\
generator by S. Kallweit.}, which contains a fully automated implementation of the dipole subtraction method for massless and massive partons
at NLO QCD~\cite{Catani:1996jh,Catani:1996vz,Catani:2002hc} and NLO EW~\cite{Kallweit:2017khh,Dittmaier:1999mb,Dittmaier:2008md,Gehrmann:2010ry,Schonherr:2017qcj}.
The cancellation of the remaining infrared singularities is achieved by using a formulation of the $q_T$ subtraction formalism~\cite{Catani:2007vq} derived from the NNLO QCD computation of heavy-quark production~\cite{Catani:2019iny,Catani:2019hip,Catani:2020kkl} through a suitable abelianisation procedure~\cite{deFlorian:2018wcj,Buonocore:2019puv}.
The paper is organised as follows. In Sec.~\ref{sec:HV-PA} we set up the formalism and detail the use of the PA and the reweighting procedure exploited in the computation. In Sec.~\ref{sec:num} we validate our method and present our results for the mixed QCD--EW corrections to the charged-current DY process. Our findings are summarised in Sec.~\ref{sec:summa}.

\section{Hard virtual function in the pole approximation}
\label{sec:HV-PA}

The differential cross section for the process in Eq.~(\ref{eq:proc}) can be written as
\begin{equation}
  \label{eq:exp}
  d{\sigma}=\sum_{m,n=0}^\infty d{\sigma}^{(m,n)}\, ,
\end{equation}
where $d{\sigma}^{(0,0)}\equiv d{\sigma}_{\rm LO}$ is the Born level contribution and $d{\sigma}^{(m,n)}$ the ${\cal O}(\as^m\alpha^n)$ correction.
The mixed QCD--EW corrections correspond to the term $m=n=1$ in this expansion.
According to the $q_T$ subtraction formalism~\cite{Catani:2007vq} $d{\sigma}^{(m,n)}$ can be evaluated as
\begin{equation}
  \label{eq:master}
  d{\sigma}^{(m,n)}={\cal H}^{(m,n)}\otimes d{\sigma}_{\rm LO}+\left[d\sigma_{\rm R}^{(m,n)}-d\sigma_{\rm CT}^{(m,n)}\right]\, .
\end{equation}
The first term in Eq.~(\ref{eq:master}) is obtained through a convolution (denoted by the symbol $\otimes$),
with respect to the longitudinal-momentum fractions $z_1$ and $z_2$ of the colliding partons,
of the perturbatively computable function ${\cal H}^{(m,n)}$ with the LO cross section $d{\sigma}_{\rm LO}$.
The second term is the {\it real} contribution $d\sigma_{\rm R}^{(m,n)}$, where the charged lepton and the corresponding neutrino
are accompanied by additional QCD and/or QED radiation that produces a recoil with finite transverse momentum $q_T$.
For $m+n=2$ such contribution can be evaluated by using the dipole subtraction formalism~\cite{Catani:1996jh,Catani:1996vz,Catani:2002hc,Kallweit:2017khh,Dittmaier:1999mb,Dittmaier:2008md,Gehrmann:2010ry,Schonherr:2017qcj}.
In the limit $q_T\to 0$ the real contribution
$d\sigma_{\rm R}^{(m,n)}$ is divergent, since the recoiling radiation becomes soft and/or collinear to the initial-state partons.
The role of the third term, the counterterm $d\sigma_{\rm CT}^{(m,n)}$, is to cancel the singular behaviour in the limit $q_T\to 0$, thereby rendering the cross section in Eq.~(\ref{eq:master}) finite.

Besides the numerous applications at NNLO QCD for the production of colourless final-state systems (see Ref.~\cite{Grazzini:2017mhc} and references therein), and for heavy-quark production~\cite{Catani:2019iny,Catani:2019hip,Catani:2020kkl}, which correspond to the case $m=2$, $n=0$, the method has been applied in Ref.~\cite{Buonocore:2019puv} to study NLO EW corrections to the Drell--Yan process, which represents the case $m=0$, $n=1$.
The structure of the coefficients ${\cal H}^{(0,1)}$ and $d\sigma_{\rm CT}^{(0,1)}$ can indeed be straightforwardly worked out from the corresponding coefficients
entering the NLO QCD computation of heavy-quark production through a suitable abelianisation procedure~\cite{Buonocore:2019puv}.
The structure of the coefficients ${\cal H}^{(1,1)}$ and $d\sigma_{\rm CT}^{(1,1)}$ can also be derived from those controlling
the NNLO QCD computation of heavy-quark production.
The initial-state soft/collinear and purely collinear contributions were already presented in Ref.~\cite{Cieri:2020ikq}.
The fact that the final state is colour neutral implies that final-state radiation is of pure QED origin.
Therefore, the purely soft contributions (one-loop diagrams with a soft photon or gluon and tree-level diagrams with a soft photon and a soft gluon)
have a much simpler structure compared to the corresponding contributions entering the
NNLO QCD computation of Refs.~\cite{Catani:2019iny,Catani:2019hip,Catani:2020kkl} and can be directly constructed from
those appearing at ${\cal O}(\alpha)$.\footnote{The only additional technical complication is the fact that the heavy quark is replaced by a much lighter lepton whose mass regularises final state QED collinear singularities.}
The only missing perturbative ingredient is therefore the ${\cal O}(\as\alpha)$ two-loop amplitude entering the coefficient ${\cal H}^{(1,1)}$ in Eq.~(\ref{eq:master}).

The coefficient ${\cal H}^{(m,n)}$ can be decomposed as
\begin{equation}
  {\cal H}^{(m,n)}=H^{(m,n)}\delta(1-z_1)\delta(1-z_2)+\delta{\cal H}^{(m,n)}\,,
\end{equation}
where the hard contribution $H^{(m,n)}$  contains the $(m+n)$-loop virtual corrections.
More precisely, we define
\begin{equation}
  \label{eq:H01def}
  H^{(0,1)}=\frac{2{\rm Re}\left({\cal M}_{\rm fin}^{(0,1)}{\cal M}^{(0,0)*}\right)}{|{\cal M}^{(0,0)}|^2}
  \end{equation}
  and
  \begin{equation}
    \label{eq:H11def}
  H^{(1,1)}=\frac{2{\rm Re}\left({\cal M}_{\rm fin}^{(1,1)}{\cal M}^{(0,0)*}\right)}{|{\cal M}^{(0,0)}|^2} \, .  
\end{equation}
In Eqs.~(\ref{eq:H01def},\ref{eq:H11def}) ${\cal M}^{(0,0)}$ is the Born amplitude, while ${\cal M}_{\rm fin}^{(0,1)}$ and ${\cal M}_{\rm fin}^{(1,1)}$ (defined through an expansion analogous to Eq.~(\ref{eq:exp})) are the finite parts of the renormalised virtual amplitudes entering the NLO EW and the mixed QCD--EW calculations, respectively. Their explicit expressions in terms of the renormalised virtual amplitudes ${\cal M}^{(1,0)}$, ${\cal M}^{(0,1)}$, ${\cal M}^{(1,1)}$ after subtraction of the infrared poles in $d=4-2\eps$ dimensions read
\begin{align}
{\cal M}^{(1,0)}_{\rm fin}=&\;{\cal M}^{(1,0)}+ \f{1}{2}\left(\frac{\as}{\pi}\right)C_F\left[\f{1}{\eps^2} + \left(\f{3}{2} + i\pi\right)\f{1}{\eps} - \f{\pi^2}{12}   \right]\amp^{(0)}\,,
\label{eq:M10}\\
{\cal M}^{(0,1)}_{\rm fin}=&\;{\cal M}^{(0,1)}+ \f{1}{2}\left(\frac{\alpha}{\pi}\right)\left\{ \left[ \f{1}{\eps^2} + \left(\f{3}{2} + i\pi\right)\f{1}{\eps} - \f{\pi^2}{12}\right] \frac{e^2_u+e^2_d}{2} -\f{2\Gamma_t}{\eps}\right\}\amp^{(0)}\,,\label{eq:M01}\\
\label{eq:M11}
\ampF^{(1,1)}=&\;{\cal M}^{(1,1)}-\left(\frac{\as}{\pi}\right)\left(\frac{\alpha}{\pi}\right)\Bigg\{\f{1}{8\eps^4}(e_u^2+e^2_d)C_F+ \f{1}{2\eps^3}C_F\left[\left(\frac{3}{2} + i\pi \right)\frac{e_u^2+e^2_d}{2}-\Gamma_t\right]\Bigg\}\amp^{(0)}\nn\\
&+ \f{1}{2\eps^2}\bigg\{\left(\frac{\alpha}{\pi}\right)\frac{e_u^2+e^2_d}{2}\ampF^{(1,0)} + C_F\left(\frac{\as}{\pi}\right) \ampF^{(0,1)} \nn\\
&\hspace{1.1cm}+ C_F\left(\frac{\as}{\pi}\right)\left(\frac{\alpha}{\pi}\right)\left[ \left(\f{7}{12}\pi^2 -\f{9}{8} -\f{3}{2}i\pi  \right)\frac{e_u^2+e^2_d}{2} + \left(\f{3}{2}+i\pi\right)\Gamma_t\right]  \amp^{(0)} \bigg\} \nn\\
    &+\f{1}{2\eps} \bigg\{\left(\frac{\alpha}{\pi}\right)\left[ \left( \f{3}{2}+i\pi \right) \frac{e_u^2+e^2_d}{2}-2\Gamma_t \right]\ampF^{(1,0)} + \left(\frac{\as}{\pi}\right)C_F\left[ \f{3}{2}+i\pi \right]\ampF^{(0,1)}\nn\\
    & \hspace{1cm}+\f{1}{8}C_F\left(\frac{\as}{\pi}\right)\left(\frac{\alpha}{\pi}\right)\left[ \left( \f{3}{2} -\pi^2 +24\zeta(3)+\f{2}{3}i\pi^3\right)\frac{e_u^2+e^2_d}{2} -\f{2}{3}\pi^2\Gamma_t \right]\amp^{(0)} \bigg\}\,.
\end{align}
In Eqs.~(\ref{eq:M01}) and (\ref{eq:M11}) the function $\Gamma_t$ is the NLO EW soft anomalous dimension, which is
an operator acting on the charge indices of the initial-state quarks and the final-state lepton. Its explicit expression can be derived from the corresponding NLO QCD expression~\cite{Catani:2014qha} through
an abelianisation procedure~\cite{Buonocore:2019puv}. For the relevant Born level process
\begin{equation}
  u(p_1)\bar{d}(p_2)\to \ell^+(p_3) \nu_{\ell}(p_4)
\end{equation}
$\Gamma_t$ takes the form
\begin{equation}
    \Gamma_t= -\f{1}{4}\bigg\{ e^2_\ell (1-i\pi) + \sum_{i=1,2}e_ie_3\ln\f{(2p_i \cdot p_3)^2}{Q^2m_\ell^2}\bigg\}\,,
\end{equation}
where $Q^2=(p_1+p_2)^2$, $m_\ell$ is the mass of the charged lepton and the electric charges are given by $e_1=e_u=2/3$, $e_2=e_{\bar d}=1/3$, $e_3=-e_{\ell^+}=-1$.

Since our calculation of the coefficient $H^{(1,1)}$ will be carried out by using the PA, in Sec.~\ref{sec:numPA} we are going to validate this approximation by studying its equivalent $H^{(0,1)}$ at ${\cal O}(\alpha)$.
We therefore define two different approximations of the $H^{(0,1)}$ coefficient,
\begin{align}
  \label{eq:H01PA}
  H^{(0,1)}_{\rm PA}&=\frac{2{\rm Re}\left({\cal M}_{\rm fin}^{(0,1)}{\cal M}^{(0,0)*}\right)_{\rm PA}}{|{\cal M}^{(0,0)}|^2}\,,\\
  \label{eq:H01PArwt}
  H^{(0,1)}_{\rm PA,rwg}&=\frac{2{\rm Re}\left({\cal M}_{\rm fin}^{(0,1)}{\cal M}^{(0,0)*}\right)_{\rm PA}}{|{\cal M}^{(0,0)}_{\rm PA}|^2}\, .
\end{align}
In the pure PA of Eq.~(\ref{eq:H01PA})
the $ H^{(0,1)}_{\rm PA}$ coefficient is simply computed by evaluating the interference of the tree-level and the one-loop amplitude in the PA. Since the $H^{(0,1)}$ coefficient is eventually multiplied with
the Born cross section $d\sigma_{\rm LO}$ (see the first term on the right-hand side of Eq.~(\ref{eq:master})), this procedure corresponds to the standard PA.
An alternative procedure is to evaluate also the denominator in Eq.~(\ref{eq:H01def}) in the PA. This approach effectively reweights the virtual--tree interference in PA with the full squared Born amplitude, and is thus labelled  ${\rm PA,rwg}$ in Eq.~(\ref{eq:H01PArwt}).

We now turn to ${\cal O}(\as\alpha)$. Since, as we will show, the PA with the reweighting procedure works rather
well for $H^{(0,1)}$, we define two approximations for the $H^{(1,1)}$ coefficient
involving different reweighting factors,
\begin{align}
\label{eq:H11PA1}
H^{(1,1)}_{{\rm PA,rwg_B}}&=H^{(1,1)}_{{\rm PA}}\times \frac{|{\cal M}^{(0,0)}|^2}{|{\cal M}^{(0,0)}_{\rm PA}|^2}=
\frac{2{\rm Re}\left({\cal M}_{\rm fin}^{(1,1)}{\cal M}^{(0,0)*}\right)_{\rm PA}}{|{\cal M}^{(0,0)}_{\rm PA}|^2}\,,\\
\label{eq:H11PA2}
H^{(1,1)}_{{\rm PA,rwg_V}}&=H^{(1,1)}_{{\rm PA}} \times \frac{H^{(0,1)}}{H^{(0,1)}_{\rm PA}}=\frac{2{\rm Re}\left({\cal M}_{\rm fin}^{(1,1)}{\cal M}^{(0,0)*}\right)_{\rm PA}}{|{\cal M}^{(0,0)}|^2}\times\frac{2{\rm Re}\left({\cal M}_{\rm fin}^{(0,1)}{\cal M}^{(0,0)*}\right)_{\phantom{\rm PA}}}{2{\rm Re}\left({\cal M}_{\rm fin}^{(0,1)}{\cal M}^{(0,0)*}\right)_{\rm PA}}\,.
\end{align}
In the first approximation, Eq.~(\ref{eq:H11PA1}), the interference of the two-loop amplitude with the Born amplitude is computed in the PA, and the same is done for the denominator.
This is analogous to what is performed in Eq.~(\ref{eq:H01PArwt}) at NLO EW. In the second approximation, the expression of the $H^{(1,1)}_{\rm PA}$ coefficient is multiplied by the ratio $H^{(0,1)}/H^{(0,1)}_{\rm PA}$, which corresponds to a reweighting with the full EW-virtual--Born interference.
This reweighting procedure can be motivated by assuming a factorised approach: if we could write $H^{(1,1)}\sim H^{(1,0)}\times H^{(0,1)}$, such additional factor would improve the $H^{(0,1)}$ coefficient with the exact
one-loop EW virtual amplitude and, therefore, with the correct Sudakov EW logarithmic contributions \cite{Denner:2000jv,Accomando:2004de} at large values of transverse momenta. In the following we will use the prescription in Eq.~(\ref{eq:H11PA2}) to define our central prediction for the ${\cal O}(\as\alpha)$ correction. The difference between the results obtained with the two approximations
can be used as an estimate of the uncertainty of our procedure.

\section{Numerical results}
\label{sec:num}

Having described our calculation and defined our approximations, we can now move on to present our results.
We consider the process $pp\to \mu^+\nu_\mu+X$ at centre-of-mass energy \mbox{$\sqrt{s}=14$\,TeV}. 
As for the EW couplings, we follow the setup of Ref.~\cite{Dittmaier:2015rxo}.
In particular, we use the $G_\mu$ scheme with $G_F=1.1663787\times 10^{-5}$\,GeV$^{-2}$ and set the \textit{on-shell} values of masses and widths to $m_{W,{\rm OS}}=80.385$\,GeV, $m_{Z, {\rm OS}}=91.1876$\,GeV, $\Gamma_{W, {\rm OS}}=2.085$\,GeV, $\Gamma_{Z, {\rm OS}}=2.4952$\,GeV. Following the default mass scheme in {\sc Recola}, those values are translated to the corresponding \textit{pole} values $m_{V}=m_{V,{\rm OS}}/\sqrt{1+\Gamma^2_{V,{\rm OS}}/m^2_{V{,\rm OS}}}$ and $\Gamma_{V}=\Gamma_{V,{\rm OS}}/\sqrt{1+\Gamma^2_{V,{\rm OS}}/m_{V,{\rm OS}}^2}$, $V=W,Z$, from which  $\alpha=\sqrt{2}\,G_F m_{W}^2(1-m_{W}^2/m_{Z}^2)/\pi$ is derived,
and we use the complex-mass scheme \cite{Denner:2005fg} throughout.
The muon mass is fixed to $m_\mu=105.658369$\,MeV, and the pole masses of the top quark and the Higgs boson to $m_t=173.07$\,GeV and $m_H=125.9$\,GeV, respectively.
The CKM matrix is taken to be diagonal. We use the \texttt{NNPDF31$\_$nnlo$\_$as$\_$0118$\_$luxqed} set of parton distributions~\cite{Bertone:2017bme}, which is based on the LUXqed methodology~\cite{Manohar:2016nzj}
for the determination of the photon flux.
The QCD coupling $\as$ is evaluated at the corresponding (3-loop) order.
The renormalisation and factorisation scales are fixed to $\mu_R=\mu_F=m_W$.
At Born level this process proceeds through quark--antiquark annihilation. NLO QCD (EW) corrections additionally involve quark--gluon (quark--photon) channels, and the mixed ${\cal O}(\as\alpha)$ corrections additionally involve the gluon--photon  channel and (anti)quark--(anti)quark interference channels. All the partonic channels are included in our calculation, although, as will be shown below,
the dominant contribution is given by the quark--antiquark and quark--gluon channels.

We use the following selection cuts,
\begin{equation}
  \label{eq:cuts}
  p_{T,\mu}>25\,{\rm GeV}\,,\qquad |y_\mu|<2.5\,,\qquad p_{T,\nu}>25\,{\rm GeV}\,,
\end{equation}
and work at the level of {\it bare muons}, i.e., no lepton recombination with close-by photons is carried out.

\subsection{Validation of the pole approximation}
\label{sec:numPA}

Before presenting our results, we study the quality of the PA for the $H^{(0,1)}$ coefficient.
\begin{figure}[t]
\begin{center}
\includegraphics[width=0.46\textwidth]{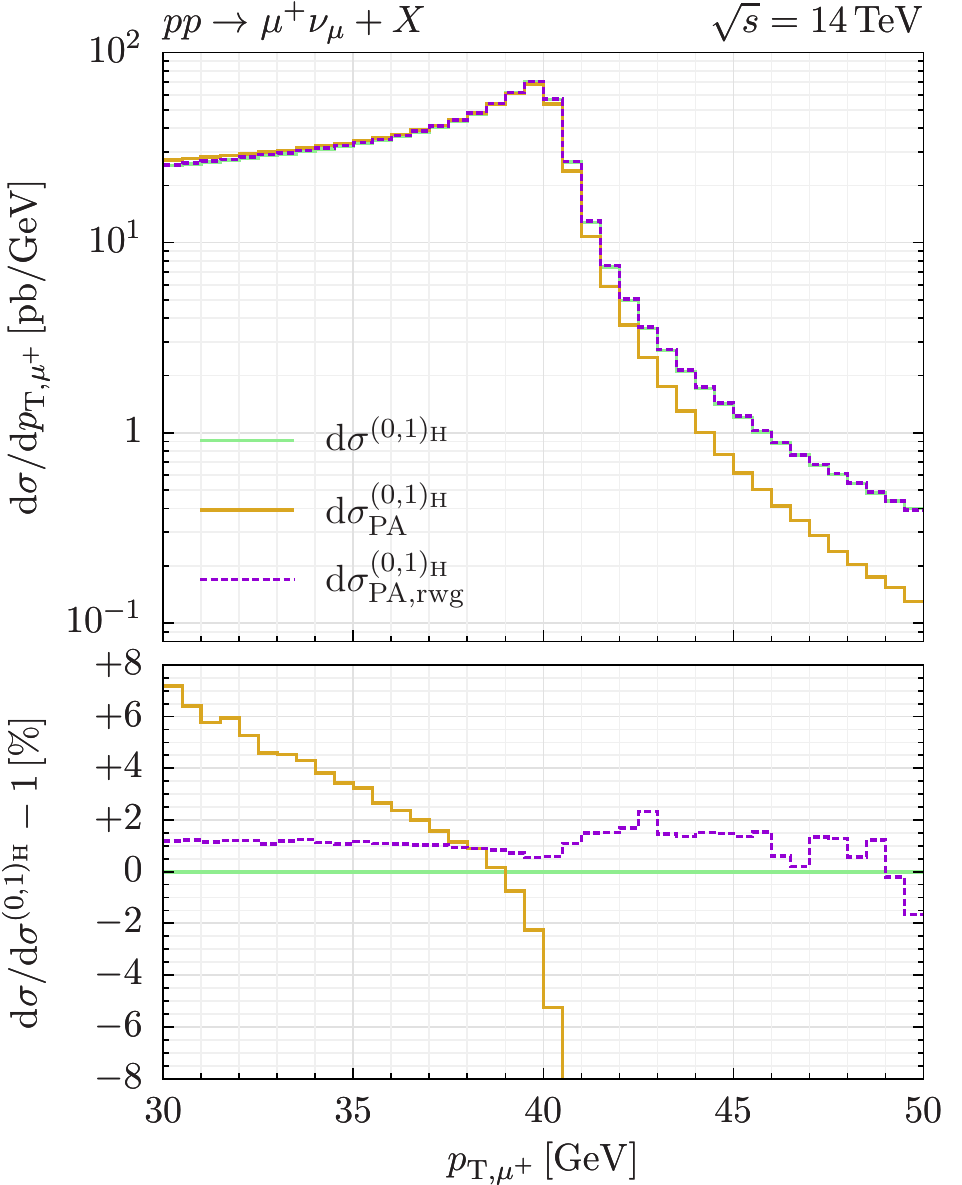}\hfill
\includegraphics[width=0.46\textwidth]{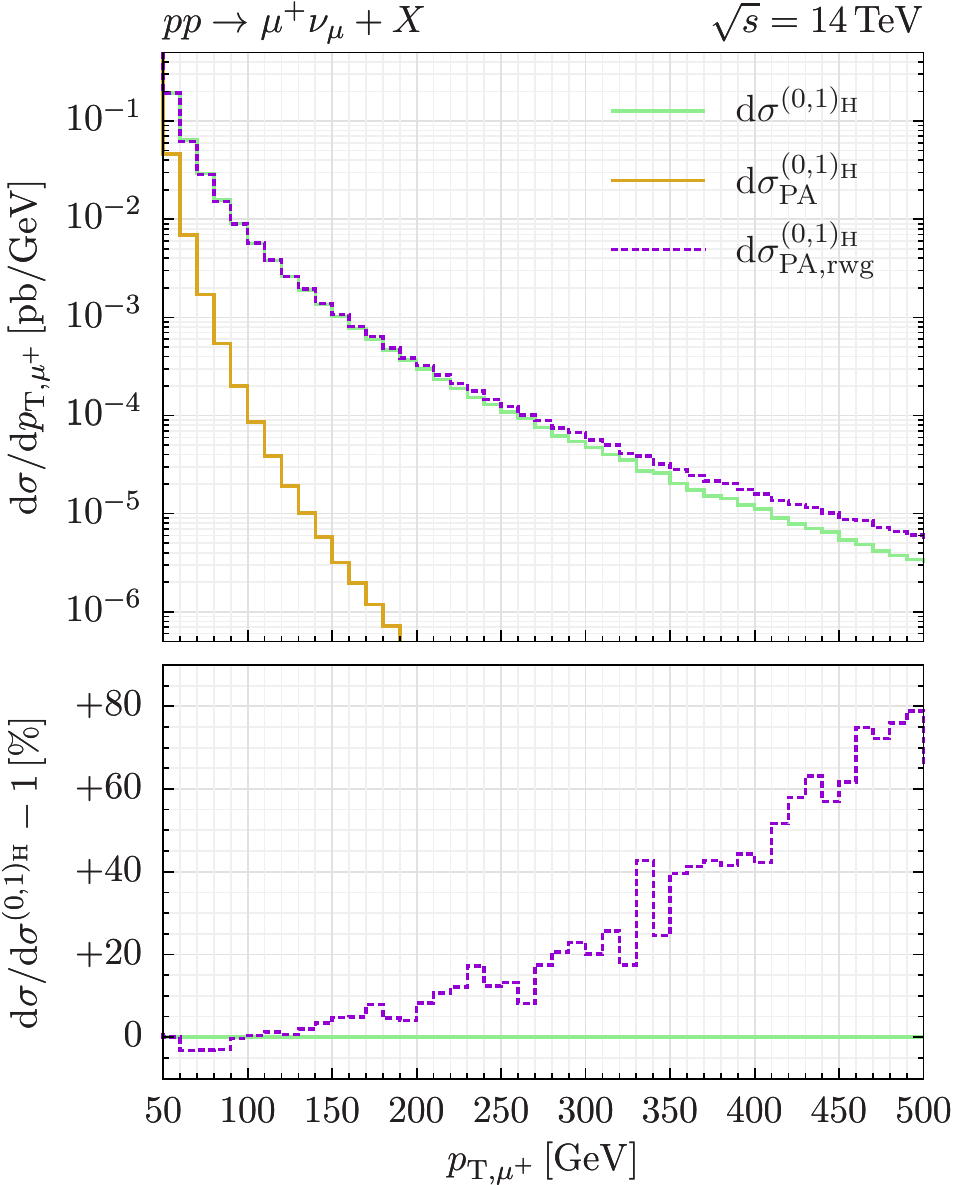}\\
\end{center}
\vspace{-2ex}
\caption{\label{fig:H1}
Contributions $d\sigma^{(0,1)_{\rm H}}_{\mathrm{PA}}$ and $d\sigma^{(0,1)_{\rm H}}_{\mathrm{PA,rwg}}$ of the hard coefficient $H^{(0,1)}$ in the approximations introduced in Eqs.~(\ref{eq:H01PA}) and (\ref{eq:H01PArwt}), respectively, compared with the exact result $d\sigma^{(0,1)_{\rm H}}$: the region of the Jacobian peak (left) and the tail of the distribution in the muon $p_T$ (right).
The upper panels show the absolute predictions, and the lower panels display the relative difference to $d\sigma^{(0,1)_{\rm H}}$.
}
\end{figure}
In Fig.~\ref{fig:H1} we show the contribution of the coefficient $H^{(0,1)}$ to the NLO EW correction as a function of the muon transverse momentum ($p_{T}$).
In particular, we compare the two approximations of Eqs.~(\ref{eq:H01PA}) and (\ref{eq:H01PArwt}) to the exact result.
The left panel depicts the region around the Jacobian peak, and the right panel the high-$p_T$ region.
In the region below the Jacobian peak the PA works relatively well and reproduces the exact result up to a few percent.
On the contrary, as $p_T$ increases, $H^{(1,0)}_{\rm PA}$ significantly undershoots the exact result.
However, when the PA is improved through the reweighting procedure, the $H^{(0,1)}$ coefficient is reproduced sufficiently well. The relative differences are at the percent level in the low-$p_T$ region and remain under control up to high transverse momenta. To which extent the approximation is good, depends, however, on the actual impact of the $H^{(0,1)}$ contribution to the complete NLO EW correction.
This is, of course, not an issue at ${\cal O}(\alpha)$ since $H^{(0,1)}$ is available.

\begin{figure}[t]
\begin{center}
\includegraphics[width=0.46\textwidth]{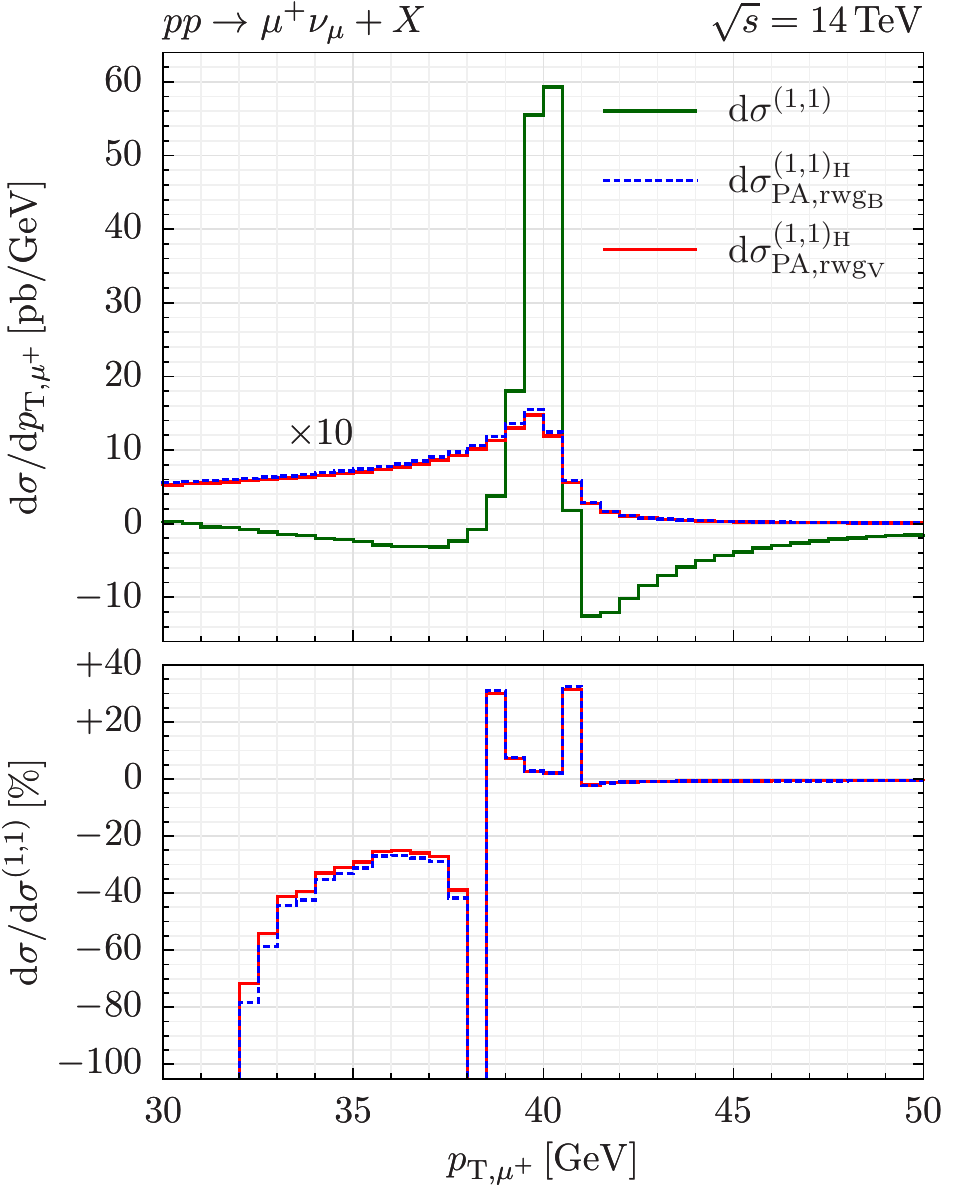}\hfill
\includegraphics[width=0.46\textwidth]{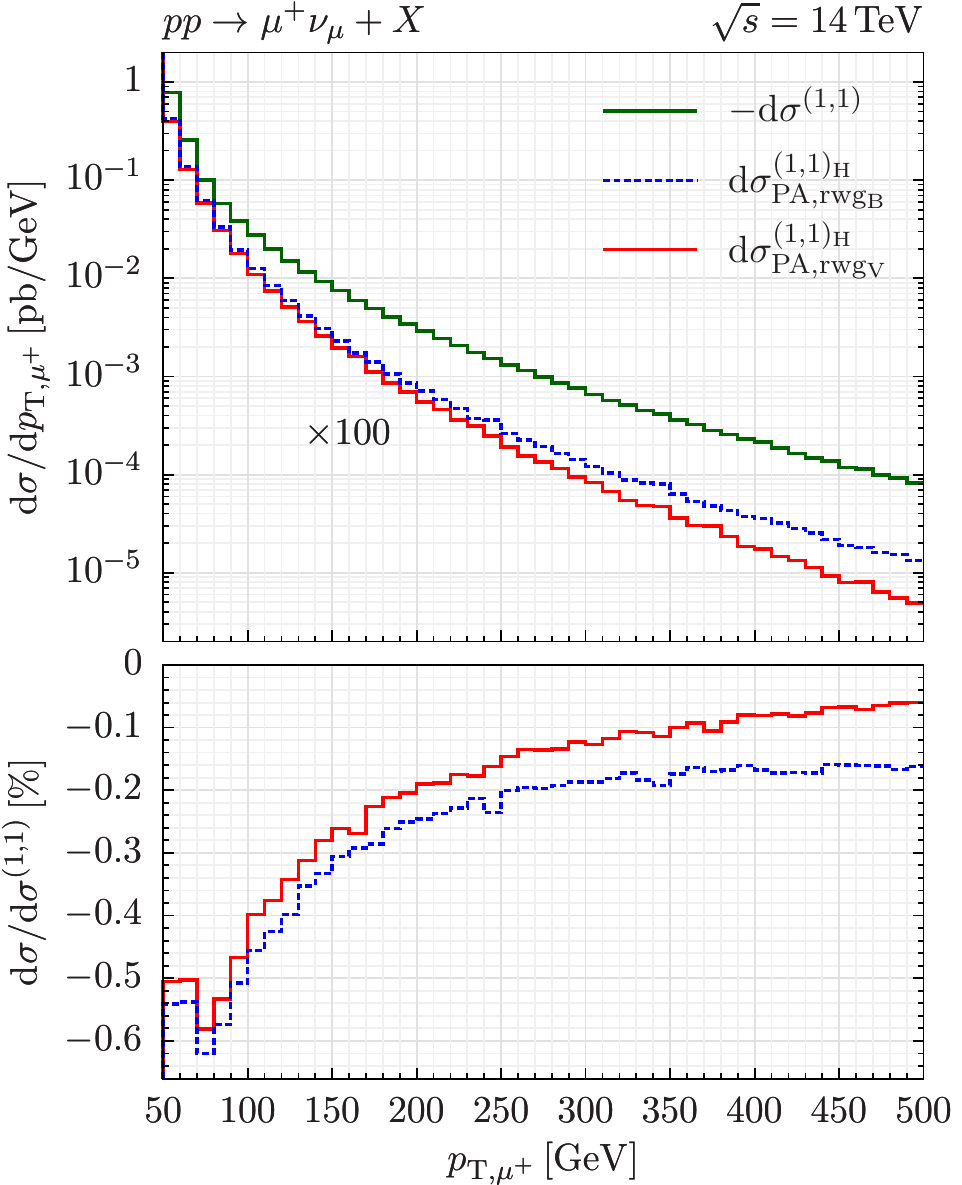}\\
\end{center}
\vspace{-2ex}
\caption{\label{fig:H11} Impact of the $\mathcal{O}(\as \alpha)$ hard-virtual contribution on the mixed QCD--EW corrections, computed adopting the reweighting procedures introduced in Eqs.~(\ref{eq:H11PA1}) and (\ref{eq:H11PA2}), respectively: the region of the Jacobian peak (left) and the tail of the distribution in the muon $p_T$ (right).
The upper panels show the absolute predictions, with the contributions $d\sigma^{(1,1)_{\rm H}}_{\mathrm{PA,rwg_B}}$ and $d\sigma^{(1,1)_{\rm H}}_{\mathrm{PA,rwg_V}}$ multiplied by a factor of 10 (100) in the peak (tail) region.
The lower panels show the effect of $d\sigma^{(1,1)_{\rm H}}_{\mathrm{PA,rwg_{B/V}}}$ normalized to $d\sigma^{(1,1)}$.}
\end{figure}
We now consider the $H^{(1,1)}$ coefficient. In Fig.~\ref{fig:H11} we show our result for the ${\cal O}(\as\alpha)$ correction, obtained by using the approximation in Eq.~(\ref{eq:H11PA2}), and the contribution
of the $H^{(1,1)}$ coefficient in the two approximations.
The lower panel displays the relative impact of the $H^{(1,1)}$ contribution.
We first discuss the low-$p_T$ region. Here the relative impact of the $H^{(1,1)}$ contribution can be quite large, being of ${\cal O}(-30\%)$ for very low $p_T$ values
and reaching ${\cal O}(+30\%)$ in the regions $p_T\sim 38$--$39$\,GeV and $p_T\sim 40$--$41$\,GeV. However, these are the $p_T$ regions where the ${\cal O}(\as\alpha)$ correction is either very small or changes sign.
On the other hand, given our observations at ${\cal O}(\alpha)$, the reweighted pole approximation is expected to work well in these regions, and the two approximations defined above are indeed in good agreement.
The situation is very different in the high-$p_T$ region. Here the impact of the $H^{(1,1)}$ coefficient is smaller than $1\%$ of the entire ${\cal O}(\as\alpha)$ correction.
We also observe that the contribution of $H^{(1,1)}$ becomes even smaller as $p_T$ increases, being
at the permille level at very large $p_T$ values.
This is not unexpected since at large $p_T$ the resonant Born-like topologies are suppressed and the cross section is dominated by real contributions, where an on-shell $W$ boson is accompanied by
hard QCD or QED emissions. More precisely, we find that in this region the dominant contribution is given by the quark--gluon partonic channel, which is computed exactly.
Therefore, even if our approximations in Eqs.(\ref{eq:H11PA1}) and (\ref{eq:H11PA2}) should fail, say, by a factor of two in this region,
the impact on the computed correction would be negligible.
Since all the other contributions in Eq.~(\ref{eq:master}) are treated exactly,
we conclude that our calculation of the complete ${\cal O}(\as\alpha)$ correction can be considered reliable both at small and large values of $p_T$.

\begin{figure}[t]
\begin{center}
\includegraphics[width=0.46\textwidth]{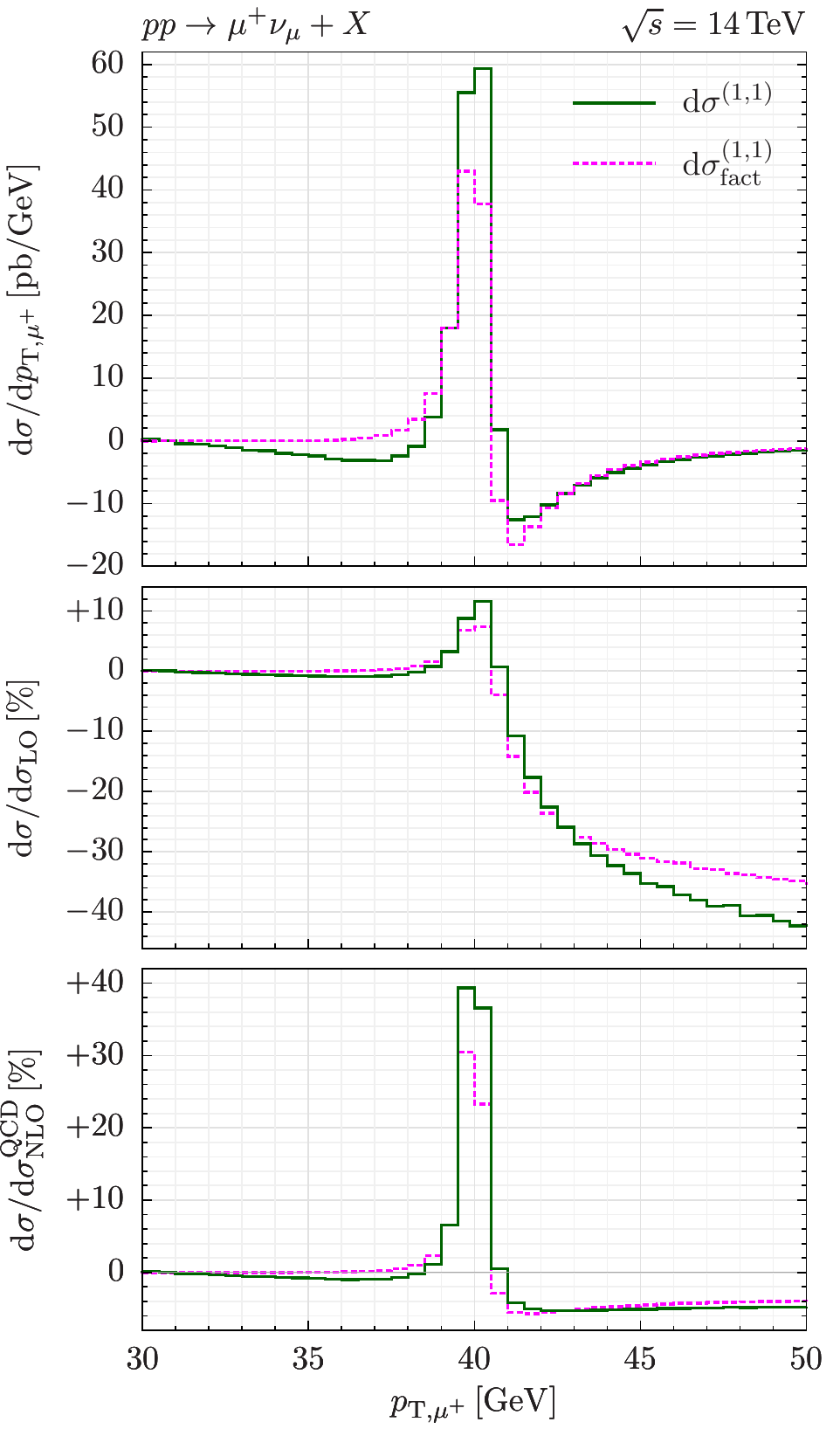}\hfill
\includegraphics[width=0.46\textwidth]{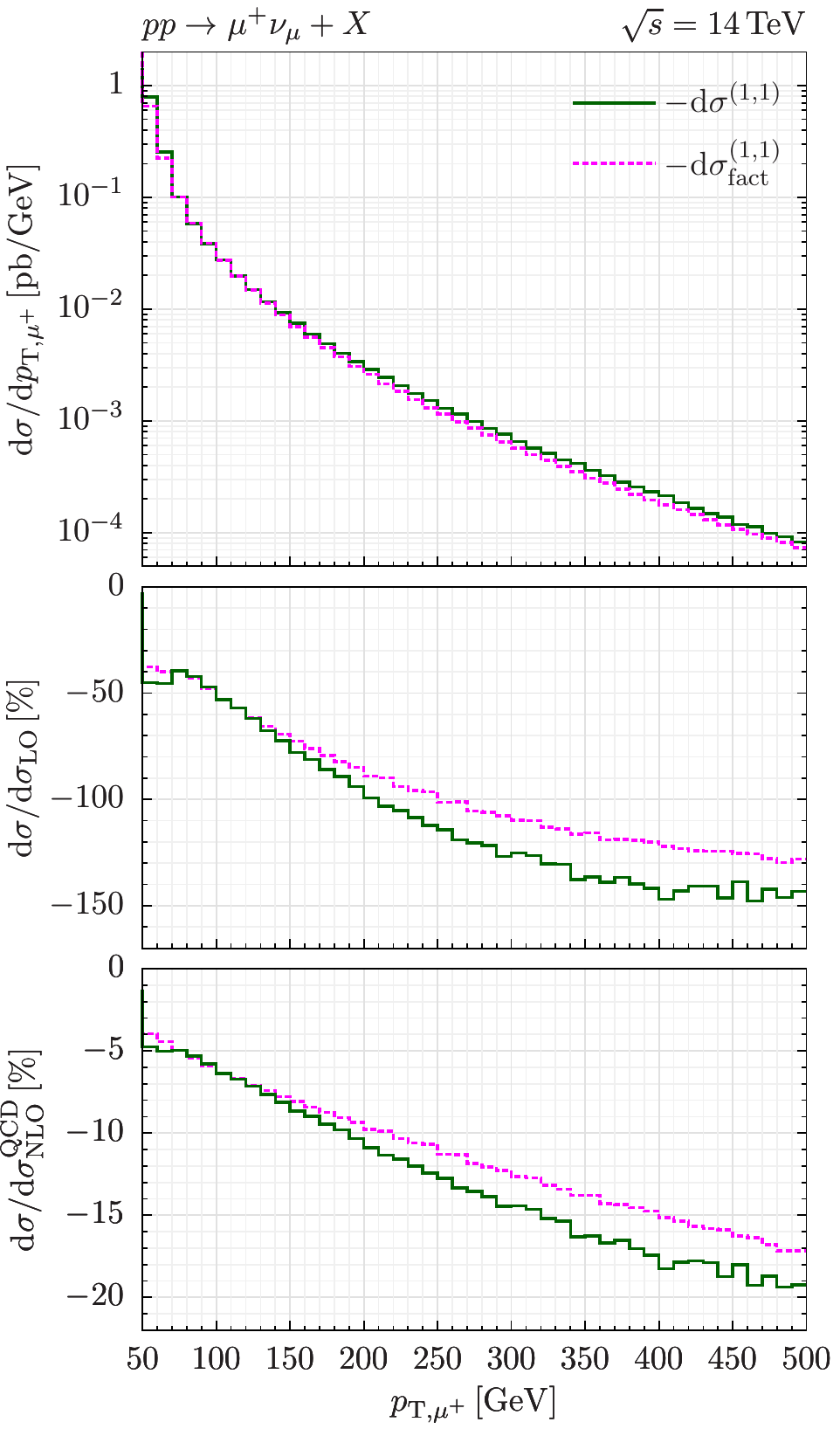}\\
\end{center}
\vspace{-2ex}
\caption{\label{fig:corr}
Complete ${\cal O}(\as\alpha)$ correction to the differential cross section $d\sigma^{(1,1)}$ in the muon $p_T$, and its factorized approximation $d\sigma^{(1,1)}_{\rm fact}$, defined in Eq.~(\ref{eq:fact}).
The top panels show the absolute predictions, while the central (bottom) panels display the ${\cal O}(\as\alpha)$ correction normalized to the LO (NLO QCD) result.
}
\end{figure}

\subsection[Results on the ${\cal O}(\as\alpha)$ correction]{Results on the $\boldsymbol{{\cal O}(\as\alpha)}$ correction}
\label{sec:numres}

We now turn to our final result for the complete ${\cal O}(\as\alpha)$ correction in Fig.~\ref{fig:corr}.
In the main panels we show the absolute correction $d\sigma^{(1,1)}/dp_T$ as a function of the muon $p_T$.
The central (bottom) panels display the correction normalised to the LO (NLO QCD) result.
As discussed above, the central value for $d\sigma^{(1,1)}$ is obtained by computing the hard coefficient $H^{(1,1)}$ as in Eq.~(\ref{eq:H11PA2}),
and the difference with the result obtained by using the prescription in Eq.~(\ref{eq:H11PA1})
is taken as an uncertainty. However, such uncertainty is so small that it is not resolved on the scale of these plots.
Our results can be compared with those from an approach in which QCD and EW corrections are assumed to completely factorise.
Such approximation can be defined as follows:
For each bin, the QCD correction, $d\sigma^{(1,0)}/dp_T$, and the EW correction restricted to the $q{\bar q}$ channel,
$d\sigma_{q{\bar q}}^{(0,1)}/dp_T$, are computed, and the factorised ${\cal O}(\as\alpha)$ correction is calculated as
\begin{equation}
  \label{eq:fact}
\frac{d\sigma^{(1,1)}_{\rm fact}}{dp_T}=\left(\frac{d\sigma^{(1,0)}}{dp_T}\right)\times\left(\frac{d\sigma_{q{\bar q}}^{(0,1)}}{dp_T}\right)\times\left(\frac{d\sigma_{\rm LO}}{dp_T}\right)^{-1}\, .
\end{equation}
This definition and, in particular, the exclusion of the photon-induced channels in the EW correction $d\sigma^{(0,1)}$ deserve some comments \cite{Kallweit:2019zez}. A factorised approach of mixed QCD--EW corrections is justified if the dominant sources of QCD and EW corrections factorise with respect to the hard $W$ production subprocess.
In phase space regions populated by the emission of a hard additional photon (jet), a factorised approach including the photon-induced contribution would effectively multiply {\it giant K-factors} \cite{Rubin:2010xp} of QCD and EW origin, and, therefore, is not expected to work.

\begin{figure}[t!]
  \centering
  \includegraphics[width=0.46\textwidth]{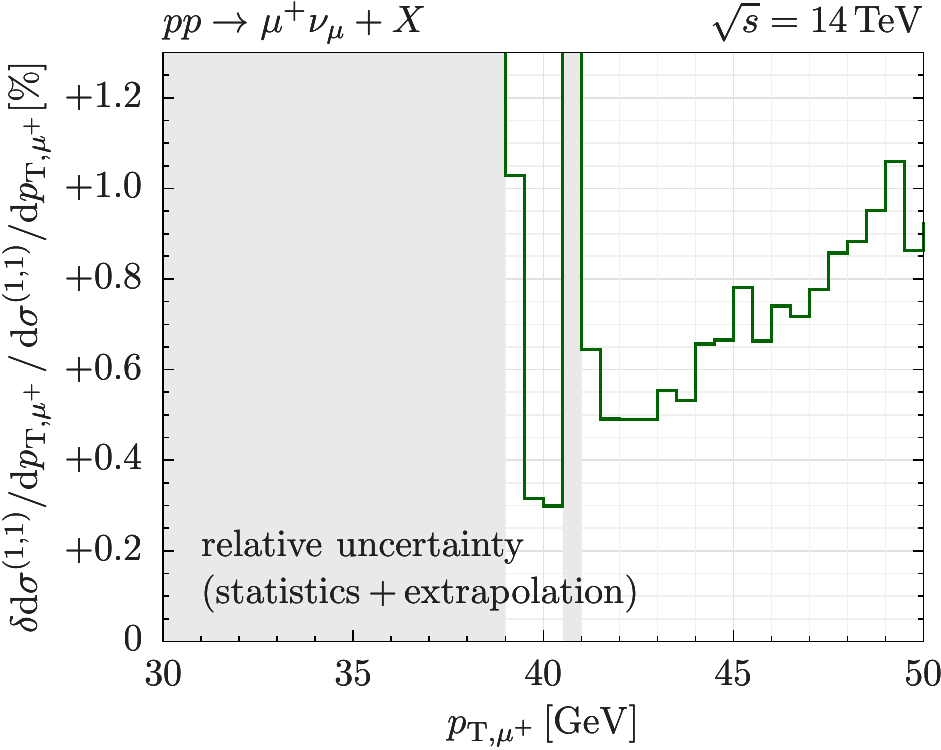}\hfill
  \includegraphics[width=0.46\textwidth]{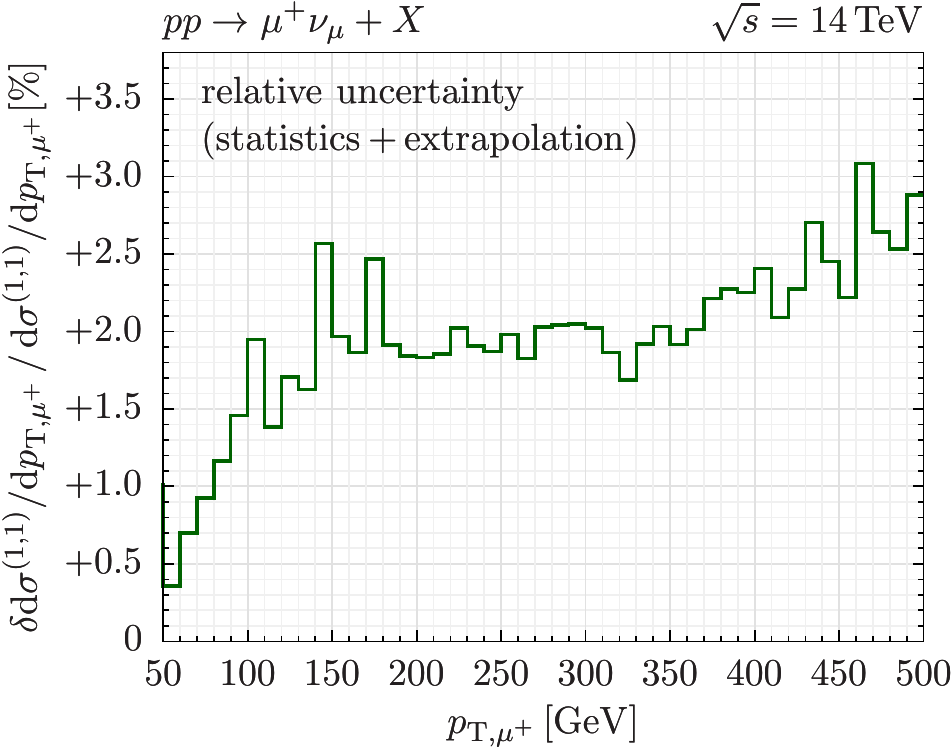}
  \caption{Relative uncertainty of the ${\cal O}(\as\alpha)$ coefficient as a function of the muon $p_T$. Combined statistical and systematic uncertainties are shown, and the latter is associated to the $q_T\to0$ extrapolation procedure described in the main text, performed on a binwise level. In the grey-shaded regions, $d\sigma^{(1,1)}/dp_T$ is approximately zero and its relative error thus meaningless.
  }\vspace*{1ex}
  \label{fig:syst}
\end{figure}

Our result for the ${\cal O}(\as\alpha)$ correction in the region of the Jacobian peak is reproduced relatively well by the factorised approximation.
Beyond the Jacobian peak, the factorised result tends to overshoot the complete result.
This is in line with what was observed in Ref.~\cite{Dittmaier:2015rxo}. As $p_T$ increases, the (negative) impact of the mixed QCD--EW corrections increases, and at $p_T=500$\,GeV it reaches about $-140\%$ with respect to the LO prediction and $-20\%$ with respect to the NLO QCD result.
The factorised approximation describes the qualitative behaviour of the complete correction relatively well, also in the tail of the distribution, but it tends to overshoot the full result as $p_T$ increases.

We add some details on the procedure used to obtain these results.
In order to  numerically evaluate the contribution in the square bracket of Eq.~(\ref{eq:master}), a technical cut-off $\rcut$ is introduced on the dimensionless variable $q_T / M$, where $M$ is the invariant mass of the lepton--neutrino system. The final result in each bin, which corresponds to the limit $\rcut \to 0$, is extracted by computing $d\sigma^{(1,1)}/dp_T$ at fixed values of $\rcut$ in the range $[0.01\%, r_{\rm max}]$. Quadratic least $\chi^2$ fits are performed for different values of $r_{\rm max}\in[0.5\%, 1\%]$. The extrapolated value is then extracted from the fit with lowest $\chi^2/$degrees-of-freedom, and the uncertainty is estimated by comparing the results obtained through the different fits. This procedure is the same as implemented in {\sc Matrix}~\cite{Grazzini:2017mhc}.  The ensuing uncertainties of the computed correction (not shown in Fig.~\ref{fig:corr}), obtained combining statistical and systematic errors, are shown in Fig.~\ref{fig:syst}. 
They range from the percent level at low $p_T$ values to ${\cal O}(3\%)$ at $p_T=500$\,GeV, with the exception of regions where $d\sigma^{(1,1)}/dp_T$
is approximately zero and thus the relative errors are artificially large. We have checked, however, that in these regions the error
is well below one permille of the respective cross section and thus phenomenologically irrelevant.

We finally present our predictions for the fiducial cross section corresponding to the selection cuts in Eq.~(\ref{eq:cuts}). In Table~\ref{tab:fid} we report
the contributions $\sigma^{(i,j)}$ to the cross section (see Eq.~(\ref{eq:exp})) in the various partonic channels. The numerical uncertainties are stated in brackets, and for the NNLO corrections
$\sigma^{(2,0)}$ and the mixed QCD--EW contributions $\sigma^{(1,1)}$ they include the systematic uncertainties from the $\rcut\to 0$ extrapolation.
The contribution from the channels $u{\bar d},\,c{\bar s}$ is denoted by $q{\bar q}$.
\begin{table}
\renewcommand{\arraystretch}{1.4}
  \centering
  \begin{tabular}{|c|c|c|c|c|c|}
    \hline
    $\sigma$ [pb] & $\sigma_{\rm LO}$  & $\sigma^{(1,0)}$  & $\sigma^{(0,1)}$  & $\sigma^{(2,0)}$ & $\sigma^{(1,1)}$ \\
    \hline
    \hline
    $q{\bar q}$ & $5029.2$ & $\phantom{+0}970.5(3)\phantom{00}$ & $-143.61(15)$ & $\phantom{+}251(4)\phantom{.0}$ & $\hspace*{-0.6ex}\phantom{0}-7.0(1.2)\phantom{00}\hspace*{-0.6ex}$\\
    \hline
    $qg$ & --- & $-1079.86(12)$ & --- & $-377(3)\phantom{.0}$ & $\hspace*{-0.6ex}\phantom{+}39.0(4)\phantom{.00}\hspace*{-0.6ex}$ \\
    \hline
    $q(g)\gamma$ & --- & --- & $\phantom{+00}2.823(1)$ & --- & $\hspace*{-0.6ex}\phantom{+0}0.055(5)\phantom{.0}\hspace*{-0.6ex}$\\
    \hline
    $q({\bar q})q^\prime$ & --- & --- & --- & $\phantom{+0}44.2(7)$ & $\hspace*{-0.6ex}\phantom{+0}1.2382(3)\phantom{.}\hspace*{-0.6ex}$\\
    \hline
    $g g$ & --- & --- & --- & $\phantom{+}100.8(8)$ & --- \\
    \hline
    \hline
    tot & $5029.2$ & $\phantom{0}$$-109.4(4)\phantom{00}$ & $-140.8(2)\phantom{00}$ & $\phantom{00}19(5)\phantom{.0}$ &  $\hspace*{-0.6ex}\phantom{+} 33.3(1.3)\phantom{00}\hspace*{-0.6ex}$ \\
    \hline
  \end{tabular}
  \caption{\label{tab:fid}
    The different perturbative contributions to the fiducial cross section (see Eq.~(\ref{eq:exp})). The breakdown into the various partonic channels is also shown.}
\renewcommand{\arraystretch}{1.0}
  \end{table}
The contributions from the channels
$qg,\,{\bar q}g$ and $q\gamma,\,{\bar q}\gamma$, which enter at NLO QCD and EW, are labelled by $qg$ and $q\gamma$, respectively.
The contribution from all the remaining quark--quark channels
$qq',\, {\bar q}{\bar q}',\, q{\bar q}'$ (excluding $u{\bar d},\,c{\bar s}$) to the NNLO QCD and mixed corrections
is labelled by $q({\bar q})q^\prime$.
Finally, the contributions from the gluon--gluon and gluon--photon channels, which are relevant only at ${\cal O}(\as^2)$ and ${\cal O}(\as\alpha)$, are denoted by $gg$ and $g\gamma$, respectively.

We see that NLO QCD corrections are subject to large cancellations between the $q{\bar q}$ and the $qg$ channels.
As a consequence, NLO QCD and  NLO EW corrections have a similar quantitative impact: the NLO QCD corrections amount to $-2.2\%$ with respect
to the LO result, while NLO EW corrections contribute $-2.8\%$. Because of similar cancellations, the NNLO QCD corrections are rather small and amount to $+0.4\%$.
The newly computed mixed QCD--EW corrections amount to $+0.6\%$ with respect to LO, and are dominated by the $qg$ channel, whereas the
photon-induced $q(g)\gamma$ channels give a negligible contribution.

As a final remark we note that the pattern of the higher-order QCD corrections to the perturbative series is strongly dependent on the choice of the renormalisation and factorisation scales. For example, the scale choice $\mu_R=\mu_F={m_W}/{2}$ leads to a more common perturbative pattern: $\sigma^{(1,0)}/\sigma_{\rm LO} = +10\%$, $\sigma^{(0,1)}/\sigma_{\rm LO} = -2.9\%$, $\sigma^{(2,0)}/\sigma_{\rm LO} =+4.2\%$, $\sigma^{(1,1)}/\sigma_{\rm LO} = +0.76\%$.

\section{Summary}
\label{sec:summa}

In this paper we have presented a new computation of the
mixed QCD--EW corrections to charged-lepton production at the LHC. The cancellation of soft and collinear singularities has been achieved by using a
formulation of the $q_T$ subtraction formalism derived from the NNLO QCD calculation for heavy-quark production through a suitable abelianisation procedure.
All the real and virtual contributions
due to initial- and final-state radiation
have been consistently included without any approximation, except for the finite part
of the two-loop virtual correction, which has been computed in the pole approximation
and suitably improved through a reweighting procedure. We have shown that our results are reliable in both on-shell and off-shell regions, thereby providing the
first prediction of the mixed QCD--EW corrections in the entire region of the lepton
transverse momentum.
Our calculation is fully differential in the momenta of the charged lepton, the corresponding neutrino, and the associated QED and QCD radiation.
Therefore, it can be used to compute arbitrary infrared-safe observables, and, in particular, we can also deal with {\it dressed leptons}, i.e.\ leptons recombined with close-by photons.
The method we have used is in principle applicable to other EW processes, as, for example, vector-boson pair production. More details on our calculation will be presented elsewhere.

\vskip 0.5cm
\noindent {\bf Acknowledgements}

\noindent

We would like to express our gratitude to Jean-Nicolas Lang and Jonas Lindert for their
continuous support on {\sc Recola} and {\sc OpenLoops}.
We wish to thank Stefano Catani and Alessandro Vicini for helpful discussions and comments on the manuscript.
This work is supported in part by the Swiss National Science Foundation (SNF) under contracts IZSAZ2$\_$173357 and 200020$\_$188464. The work of SK is supported by the ERC Starting Grant 714788 REINVENT. FT acknowledges support from INFN.

\bibliography{biblio}

\end{document}